\documentclass[preprint,prc,aps,showpacs,showkeys,groupedaddress,floatfix]{revtex4}
\usepackage{graphics}
\usepackage{graphicx}
\usepackage{epsfig}
\usepackage{amssymb}
\usepackage{amsmath}

\begin{document}
\title{Exact analytical solution to the relativistic Klein-Gordon equation with
non-central equal scalar and vector potentials}
\author{Fevziye Yasuk, Aysen Durmus and Ismail Boztosun}
\affiliation{\it Department of Physics, Erciyes University, 38039
Kayseri Turkey}
\begin{abstract}
We present an alternative and simple method for the exact solution
of the Klein-Gordon equation in the presence of the non-central
equal scalar and vector potentials by using Nikiforov-Uvarov (NU)
method. The exact bound state energy eigenvalues and corresponding
eigenfunctions are obtained for a particle bound in a potential of
$V(r,\theta ) = \frac{\alpha }{r} + \frac{\beta }{r^2\sin ^2\theta }
+ \gamma \frac{\cos \theta }{r^2\sin ^2\theta }$ type.
\end{abstract}
\maketitle
\section{Introduction}\label{sec1}
In nuclear and high energy physics, one of the interesting problems
is to obtain exact solutions of the Klein-Gordon,
Duffin-Kemmer-Petiau and Dirac equations for mixed vector and scalar
potentials. The Klein-Gordon, Dirac and Duffin-Kemmer-Petiau wave
equations are frequently used to describe the particle dynamics in
relativistic quantum mechanics. In recent years, a great deal of
efforts have been spent to solve these relativistic wave equations
for various potentials by using different methods. These
relativistic equations contain two objects: the four-vector linear
momentum operator and the scalar rest mass. They allow us to
introduce two types of potential coupling which are the four vector
potential ($V$) and the space-time scalar potential($S$).

The Klein-Gordon equation with the vector and scalar potentials can
be written as follows:
\begin{equation}
\label{eq1}
\left[ { - \left( {i\frac{\partial }{\partial t} - V(\vec {r})} \right)^2 -
\vec {\nabla }^2 + \left( {S(\vec {r}) + M} \right)^2} \right]\psi (\vec
{r}) = 0
\end{equation}
For the case $S(\vec {r}) = \pm V(\vec {r})$, the solution of the
Klein-Gordon equation has been studied recently \cite{1,2}. The
exact solutions of these equations are possible only for certain
potentials such as Coulomb, Morse, P\"{o}schl-Teller, Hulthen and
harmonic oscillator \emph{etc.} by using different methods \cite{3}.
The other exactly solvable ones are the ring-shaped potentials
introduced by Hartmann \cite{4} and Quesne \cite{5}. These
potentials involve an attractive Coulomb potential with a repulsive
inverse square potential one. In particular, the Coulombic
ring-shaped potential  \cite{6} revived in quantum chemistry by
Hartmann and coworkers \cite{7} and the oscillatory ring-shaped
potential, systematically studied by Quesne \cite{5}, have been
investigated from a quantum mechanical view-point by using various
approaches. The special case of the potential in spherical
coordinates is
\begin{equation}
\label{eq2}
V(r,\theta ) = \frac{\alpha }{r} + \frac{\beta }{r^2\sin ^2\theta } + \gamma
\frac{\cos \theta }{r^2\sin ^2\theta }
\end{equation}
introduced by Makarov \emph{et al.}  \cite{8}. This potential can be
used in quantum chemistry and nuclear physics to describe the
ring-shaped molecules like benzene and the interactions between the
deformed pairs of the nuclei.

In this paper, we introduce an alternative and simple method for the
exact solution of the Klein-Gordon equation for the case where
$S(\vec {r}) = \pm V(\vec {r})$, considering a general angle
dependent (non-central) potential by using Nikiforov-Uvarov (NU)
method \cite{9}. This method is based on solving the second-order
linear differential equations by reducing to a generalized equation
of hypergeometric type.

NU-method is used to solve Schr\"{o}dinger, Dirac, Klein-Gordon and
Duffin-Kemmer-Petiau wave equations in the presence of the
exponential type potentials such as Woods-Saxon, P\"{o}schl-Teller
\cite{10} and Hulthen  \cite{11,12} and non-central potential
\cite{13}. The aim of this study is to show that the
Nikiforov-Uvarov method can be used to obtain exact solutions of
non-central potentials for Klein-Gordon equation. Thus, radial and
angular parts of the Klein-Gordon equation with non-central
potential are solved by NU-method and it is seen that this method is
applicable to non-central type potential for relativistic wave
equations.

In the following section, the Klein-Gordon equation with equal
scalar and vector potentials. In section \ref{sec3}, the
Klein-Gordon equation in spherical coordinates for a particle in the
presence of non-central potential is separated into radial and
angular parts. Section  \ref{sec4} is devoted to a brief description
of the Nikiforov-Uvarov method. The solutions of the radial and
angular parts of the Klein-Gordon equation by using the
Nikiforov-Uvarov method in section \ref{sec5}. Finally, concluding
remarks are given in section \ref{sec6}.
\section{Klein-Gordon equation with equal scalar and vector
potentials}\label{sec2}
For the time-independent potentials we can write the total wave
function as $\psi (\vec {r},t) = e^{ - i\varepsilon t}\psi (\vec
{r})$, where $\varepsilon $ is the relativistic energy. The
three-dimensional Klein-Gordon equation with the mixed vector and
scalar potentials can be written as follows:
\begin{equation}
\label{eq3}
\left[ {\vec {\nabla }^2 + \left( {V(\vec {r}) - \varepsilon } \right)^2 -
\left( {S(\vec {r}) + M} \right)^2} \right]\psi (\vec {r}) = 0,
\end{equation}
where M is the mass, $\varepsilon $ is the energy and $S(\vec {r})$
and $V(\vec {r})$ is the scalar and vectorial potentials
respectively. Now, if we take $S(\vec {r}) = \pm V(\vec {r})$, the
Klein-Gordon equation becomes:
\begin{equation}
\label{eq4}
\left[ {\vec {\nabla }^2 - 2\left( {\varepsilon \pm M} \right)V(\vec {r}) +
\varepsilon ^2 - M^2} \right]\psi (\vec {r}) = 0
\end{equation}
This equation describes a scalar particle, \emph{i.e}, spin-0
particle. It is the Schr\"{o}dinger equation for the potential $2V$
in the non-relativistic limit. Thus, Alhaidari \emph{et al.}
concludes that only the choice $S=+V$ produces a nontrivial
nonrelativistic limit with a potential function $2V$, and not $V$.
Accordinly, it would be natural to scale the potential terms in
Eq.(\ref{eq3}) so that in the relativistic limit the interaction
potential becomes $V$, not $2V$. Therefore, they modify
Eq.(\ref{eq3}) to read as follows \cite{1}:
\begin{equation}
\label{eq5}
\left[ {\vec {\nabla }^2 + \left( {\frac{1}{2}V(\vec {r}) - \varepsilon }
\right)^2 - \left( {\frac{1}{2}S(\vec {r}) + M} \right)^2} \right]\psi (\vec
{r}) = 0
\end{equation}
Thus, Eq.(\ref{eq4}) is acquired as,
\begin{equation}
\label{eq6}
\left[ {\vec {\nabla }^2 - \left( {\varepsilon \pm M} \right)V(\vec {r}) +
\varepsilon ^2 - M^2} \right]\psi (\vec {r}) = 0
\end{equation}
In the following section, for $S(\vec {r}) = + V(\vec {r})\,$, if we
take $V(\vec {r})\,$ as a general non-central potential, three
dimensional Klein-Gordon equation is separated into variables and
these equation can be solved by using Nikiforov-Uvarov method.
\section{Separating variables of the Klein-Gordon equation with
non-central potential}\label{sec3} In the spherical coordinates, the
Klein-Gordon equation for a particle in the existence of a general
non-central potential $V(r,\theta )$ becomes
\begin{equation}
\label{eq7}
\begin{array}{l}
 \left[ {\frac{1}{r^2}\frac{\partial }{\partial r}\left( {r^2\frac{\partial
}{\partial r}} \right) + \frac{1}{r^2\sin \theta }\frac{\partial
}{\partial \theta }\left( {\sin \theta \frac{\partial }{\partial
\theta }} \right) + \frac{1}{r^2\sin ^2\theta }\frac{\partial
^2}{\partial \varphi ^2} - \left( {\varepsilon + M}
\right)V(r,\theta ) + \varepsilon ^2 - M^2} \right]\psi (r,\theta
,\varphi ) = 0
 \end{array}
\end{equation}
where $V(r,\theta )$ is a general non-central potential as given by
Eq.(\ref{eq2}). If one assigns the corresponding spherical total
wave function as $\psi (r,\theta ,\varphi ) =
\frac{1}{r}R(r)Y(\theta ,\varphi )$, then by selecting $Y(\theta
,\varphi ) = \Theta (\theta )\Phi (\varphi )$, the wave equation
(\ref{eq7}) for a general non-central potential is separated into
variables and the following equations are obtained:
\begin{equation}
\label{eq8}
\left[ {\left. {\frac{d^2}{dr^2} - \frac{\lambda }{r^2} - \left(
{\varepsilon + M} \right)\frac{\alpha }{r} + \varepsilon ^2 - M^2}
\right]R(r) = 0} \right.,
\end{equation}
\begin{equation}
\label{eq9} \frac{d^2\Theta (\theta )}{d\theta ^2} + \cot \theta
\frac{d\Theta (\theta )}{d\theta } + \left[ {\lambda -
\frac{m^2}{\sin ^2\theta } - \left( {\varepsilon + M} \right)\left(
{\frac{\beta + \gamma \cos \theta }{\sin ^2\theta }} \right)}
\right]{\kern 1pt} {\kern 1pt} \Theta (\theta ) = 0
\end{equation}
\begin{equation}
\label{eq10}
\frac{d^2\Phi (\varphi )}{d\varphi ^2} + m^2\Phi (\varphi ) = 0
\end{equation}
where m$^{2}$ and $\lambda $ are the separation constants. The
solution of equation (\ref{eq10}) is well-known and it is the
azimuthal angle solution,
\begin{equation}
\Phi _m = Ae^{im\varphi } \quad ,(m=0, \pm 1, \pm 2 \ldots )
\end{equation}
Eqs. (\ref{eq8}) and (\ref{eq9}) are radial and polar-angle
equations and they will be solved by using Nikiforov-Uvarov method
\cite{9}, given briefly in the following section.
\section{Nikiforov-Uvarov Method}\label{sec4}
The non-relativistic Schr\"{o}dinger equation or similar
time-independent second-order differential equations can be solved
by using Nikiforov-Uvarov method which is based on the solutions of
a general second-order linear differential equation with special
orthogonal functions. In this method, for a given real or complex
potential, the Schr\"{o}dinger equation is transformed into a
generalized equation of hypergeometric type with an appropriate $s =
s(r)$ coordinate transformation and it can be written in the
following form
\begin{equation}
\label{eq11}
\psi (s{)}'' + \frac{\tilde {\tau }(s)}{\sigma (s)}{\psi }'(s) +
\frac{\tilde {\sigma }(s)}{\sigma ^2(s)}\psi (s) = 0
\end{equation}
where $\sigma (s)$ and $\tilde {\sigma }(s)$ are polynomials, at
most second-degree, and $\tilde {\tau }(s)$ is a first-degree
polynomial. Hence, from Eq. (\ref{eq11}), the Schr\"{o}dinger
equation or the Schr\"{o}dinger-like equations can be solved by
means of this method for potentials we consider. In order to find a
particular solution of Eq. (\ref{eq11}), we use the separation of
variables with the transformation
\begin{equation}
\label{eq12}
\psi (s) = \phi (s)y(s)
\end{equation}
it reduces Eq. (\ref{eq11}) to an equation of hypergeometric type,
\begin{equation}
\label{eq13} \sigma (s){y}'' + \tau (s){y}' + \lambdabar y = 0
\end{equation}
and $\phi (s)$ is defined as a logarithmic derivative in the
following form and its solutions can be obtained from
\begin{equation}
\label{eq14}
{{\phi }'(s)} \mathord{\left/ {\vphantom {{{\phi }'(s)} \phi }} \right.
\kern-\nulldelimiterspace} \phi (s) = {\pi (s)} \mathord{\left/ {\vphantom
{{\pi (s)} {\sigma (s)}}} \right. \kern-\nulldelimiterspace} {\sigma (s)}
\end{equation}
The other part $y(s)$ is the hypergeometric type function whose
polynomial solutions are given by Rodrigues relation
\begin{equation}
\label{eq15}
y_n (s) = \frac{B_n }{\rho (s)}\frac{d^n}{ds^n}\left[ {\sigma ^n(s)\rho (s)}
\right]
\end{equation}
where $B_n $ is a normalizing constant and the weight function $\rho
(s)$ must satisfy the condition
\begin{equation}
\label{eq16}
\left( {\sigma {\kern 1pt} \rho } \right)^\prime = \tau {\kern 1pt} \rho
\end{equation}
The function $\pi $ and the parameter $\lambdabar $ required for
this method are defined as follows
\begin{equation}
\label{eq17}
\pi (s) = \frac{{\sigma }' - \tilde {\tau }}{2}\pm \sqrt {\left(
{\frac{{\sigma }' - \tilde {\tau }}{2}} \right)^2 - \tilde {\sigma } +
k\sigma } \quad ,
\end{equation}
\begin{equation}
\label{eq18} \lambdabar = k + {\pi }'
\end{equation}
On the other hand, in order to find the value of $k$, the expression
under the square root must be square of a polynomial. Thus, a new
eigenvalue equation for the Schr\"{o}dinger equation becomes
\begin{equation}
\label{eq19} \lambdabar = \lambdabar_n = - n{\tau }' - \frac{n(n -
1)}{2}{\sigma }''
\end{equation}
where
\begin{equation}
\label{eq20}
\tau (s) = \tilde {\tau }(s) + 2\pi (s)
\end{equation}
and its derivative is negative. By comparison of Eqs. (\ref{eq18})
and (\ref{eq19}), we obtain the energy eigenvalues.
\section{Solutions of the Radial and Angle-Dependent equations}\label{sec5}
\subsection{Solutions of the Radial Equation and Energy Eigenvalues}
The radial part of the Klein Gordon equation is given as
\begin{equation}
\label{eq21}
\left[ {\left. {\frac{d^2}{dr^2} - \frac{\lambda }{r^2} - \left(
{\varepsilon + M} \right)\frac{\alpha }{r} + \varepsilon ^2 - M^2}
\right]R(r) = 0} \right.
\end{equation}
This equation can be further arranged as
\begin{equation}
\label{eq22} {R}''(r) + \left( { - \eta ^2r^2 - \xi ^2r - \lambda }
\right)\frac{1}{r^2}R(r) = 0
\end{equation}
with
\begin{equation}
\varepsilon ^2 - M^2 = - \eta ^2, \quad \left( {\varepsilon + M}
\right) = \xi ^2,  \quad \lambda = \ell (\ell + 1), \quad \alpha = -
Ze^2
\end{equation}
which is now amenable to a NU solution. In order to find the
solution of this equation, it is necessary to compare
Eq.(\ref{eq22}) with Eq.(\ref{eq11}). By comparison, we obtain the
following polynomials:
\begin{equation}
\tilde {\tau } = 0, \quad  \quad \sigma = r, \quad  \quad \tilde
{\sigma } = - \eta ^2r^2 - \xi ^2r - \lambda
\end{equation}
Substituting these polynomials in Eq.(\ref{eq17}), we obtain $\pi $
function as
\begin{equation}
\label{eq23} \pi = \frac{1}{2}\pm \frac{1}{2}\sqrt {4\eta ^2r^2 +
4r(k + \xi ^2) + 4\lambda + 1}
\end{equation}
The expression in the square root must be square of polynomial in
respect of the NU method.

Therefore, we can determine the constant $k$ by using the condition
that the discriminant of the square root is zero, that is
\begin{equation}
\label{eq24} k = - \xi ^2\pm 2\sqrt {\eta ^2} \left( {\ell +
\frac{1}{2}} \right)
\end{equation}
In view of that, one can find new possible functions for each $k$ as
\begin{equation}
\label{eq25}
\pi = \left\{ {{\begin{array}{*{20}c}
 {\begin{array}{l}
 \frac{1}{2}\pm \left[ {\sqrt {\eta ^2} r + \left( {\ell +
\frac{1}{2}} \right)} \right],{\begin{array}{*{20}c}  \hfill &
{\kern 1pt} \mbox{for} \hfill & {k = - \xi ^2
+ 2\sqrt {\eta ^2} \left( {\ell + \frac{1}{2}} \right)} \hfill \\
\end{array} } \\
 \\
 \end{array}} \hfill \\
 {\frac{1}{2}\pm \left[ {\sqrt {\eta ^2} r - \left( {\ell +
\frac{1}{2}} \right)} \right],{\begin{array}{*{20}c}
 \hfill & \mbox{for}
\hfill & {k = - \xi ^2 - 2\sqrt {\eta ^2} \left( {\ell +
\frac{1}{2}}
\right)} \hfill \\
\end{array} }} \hfill \\
\end{array} }} \right.
\end{equation}
For the polynomial of $\tau = \tilde {\tau } + 2\pi $ which has a
negative derivative, we get
\begin{equation}
{\kern 1pt} {\kern 1pt} k = - \xi ^2 - 2\sqrt {\eta ^2} \left( {\ell
+ \frac{1}{2}} \right) \quad \mbox{and} \quad \pi = \frac{1}{2} -
\left[ {\sqrt {\eta ^2} r - \left( {\ell + \frac{1}{2}} \right)}
\right]
\end{equation}
Using $\lambdabar = k + {\pi }'$ together with the values $k$ and
$\pi$, $\tau $ and $\lambdabar $ can be respectively obtained as
\begin{equation}
\label{eq26} \tau = 2\left( {\ell + 1 - \sqrt {\eta ^2} r} \right)
\end{equation}
\begin{equation}
\label{eq27} \lambdabar = - \xi ^2 - \sqrt {\eta ^2} \left( {2\ell +
2} \right)
\end{equation}
Another definition of $\lambdabar_N$ is given at Eq.(\ref{eq19}),
\begin{equation}
\label{eq28} \lambdabar_N = 2N\sqrt {\eta ^2}
\end{equation}
comparing this with Eq.(\ref{eq27}) and inserting the values of
$\eta$ and $\xi$, the exact energy eigenvalues of radial part of
Klein-Gordon equation with non-central potential are derived as
\begin{equation}
\label{eq29}
\varepsilon _{N\ell } = M\frac{\left( {\left( {N + \ell + 1} \right)^2 -
\frac{\alpha ^2}{4}} \right)}{\left( {(N + \ell + 1)^2 + \frac{\alpha
^2}{4}} \right)}
\end{equation}
where $N$ denotes the radial quantum number. This is not equal to
the well-known positive energy spectrum of the relativistic
Klein-Gordon-Coulomb problem but gives the correct non-relativistic
limit in the case of weak coupling.

Using $\sigma $ and $\pi $ in Eqs.(\ref{eq13}) to (\ref{eq15}), we
can find the wave functions $y(r) = y_{N\ell } (r)$ and $\phi (r)$:
\begin{equation}
\label{eq30}
R_{N\ell } (z) = C_{N\ell } z^{\ell + 1}\exp ( - \frac{z}{2})L_N^{2\ell + 1}
(z)
\end{equation}
where $L_N^{2\ell + 1} (z)$ stands for the associated Laguerre
functions whose argument is equal to $z = \frac{\left( {\varepsilon
+ M} \right)Ze^2}{(N + \ell + 1)}r$ and $C_{N\ell } $ is
normalization constant determined by $\int\limits_0^\infty {R_{N\ell
}^2 } (r)dr = 1$ \cite{14}, the corresponding normalized wave
functions are finally obtained as
\begin{equation}
\label{eq31}
\begin{array}{l}
 R_{{n}'\ell } (r) = \left( {\frac{2(\varepsilon + M)Ze^2}{{n}'}} \right)^{1
\mathord{\left/ {\vphantom {1 2}} \right. \kern-\nulldelimiterspace}
2}\left( {\frac{({n}' - \ell - 1)!}{{n}'{\kern 1pt} {\kern 1pt} \Gamma
\left( {{n}' + \ell + 1} \right)\,}} \right)^{1 \mathord{\left/ {\vphantom
{1 2}} \right. \kern-\nulldelimiterspace} 2}{\kern 1pt} \left(
{\frac{2(\varepsilon + M)Ze^2}{{n}'}} \right)^{\ell + 1}\times \\
 \,\,\,\,\,\,\,\,\,\,\,\,\,\,\,\,\,\,\,\,\times r^{\ell + 1}\exp \left( { -
\frac{(\varepsilon + M)Ze^2}{{n}'}r} \right)L_{{n}' - \ell - 1}^{2\ell + 1}
\left( {\frac{(\varepsilon + M)Ze^2}{{n}'}r} \right) \\
 \end{array}
\end{equation}
where ${n}' = N + \ell + 1$. This equation is also stands for
solution of the radial Klein-Gordon equation with Coulomb potential,
since radial Klein-Gordon equation with non-central potential
contains only Coulombic potential terms.
\subsection{Eigenvalues and Eigenfunctions of the Angle-dependent
Equation} As for the solutions of angle-dependent part of the
Klein-Gordon equation, we may also derive eigenvalues and
eigenfunctions of polar angle part of the Klein-Gordon equation
similar to the method as given in section \ref{sec4}.

Eq.(\ref{eq9}) can be written in the following form by introducing a
new variable , $x = \cos \theta $,
\begin{equation}
\label{eq32}
\frac{d^2\Theta (x)}{dx^2} - \frac{2x}{1 - x^2}\frac{d\Theta (x)}{dx} +
\left( {\frac{\lambda (1 - x^2) - m^2 - (\varepsilon + M)\left( {\beta +
\gamma x} \right)}{(1 - x^2)^2}} \right)\Theta (x) = 0
\end{equation}
To apply the Nikiforov-Uvarov method, we compare Eq.(\ref{eq32})
with Eq.(\ref{eq11}). By comparison, we obtain the following
polynomials
\begin{equation} \tilde {\tau } = - 2x, \quad \sigma = 1
- x^2,\quad \tilde {\sigma } = - \lambda x^2 - \gamma x + \left(
{\lambda - m^2 - \beta } \right)
\end{equation}
The function $\pi $ is obtained by putting the above-expression in
Eq.(\ref{eq17}),
\begin{equation}
\pi = \pm \sqrt {x^2(\lambda - k) + \gamma x - (\lambda - m^2 -
\beta - k)}
\end{equation}
The expression in the square root must be square of a polynomial.
Then, one can find new possible functions for each $k$ as
\begin{equation}
\label{eq33}
\pi = \pm \left\{ {{\begin{array}{*{20}c}
 {\begin{array}{l}
 x\sqrt {\frac{m^2 + \beta + u}{2}} + \sqrt {\frac{m^2 + \beta -
 u}{2}},
{\begin{array}{*{20}c}
 \hfill & \mbox{for} \hfill & {k
= \frac{2\lambda - m^2 - \beta }{2} - \frac{1}{2}u} \hfill \\
\end{array} } \\
 \\
 \end{array}} \hfill \\
 {x\sqrt {\frac{m^2 + \beta - u}{2}} + \sqrt {\frac{m^2 + \beta +
 u}{2}},
{\begin{array}{*{20}c}
 \hfill & \mbox{for}
\hfill & {k = \frac{2\lambda - m^2 - \beta }{2} + \frac{1}{2}u} \hfill \\
\end{array} }} \hfill \\
\end{array} }} \right.
\end{equation}
where $u = \sqrt {(m^2 + \beta )^2 - \gamma ^2} $. For the
polynomial of $\tau = \tilde {\tau } + 2\pi $ which has a negative
derivative,
\begin{equation}
\label{eq34}
\tau = - 2\sqrt {\frac{m^2 + \beta - u}{2}} - 2x\left( {1 + \sqrt {\frac{m^2
+ \beta + u}{2}} } \right)
\end{equation}
Using $\lambdabar = k + {\pi }'$ and its other definition
$\lambdabar_n = - n{\tau }' - \frac{n(n - 1)}{2}{\sigma }''$ given
by Eqs. (\ref{eq18}) and (\ref{eq19}), following expressions for the
$\lambdabar$ are obtained respectively
\begin{equation}
\label{eq35} \lambdabar = \frac{2\lambda - (m^2 + \beta )}{2} -
\frac{1}{2}u - \sqrt {\frac{m^2 + \beta + u}{2}}
\end{equation}
\begin{equation}
\label{eq36} \lambdabar_n = 2n\left( {1 + \sqrt {\frac{m^2 + \beta +
u}{2}} } \right) + n(n - 1)
\end{equation}
Equating Eqs.(\ref{eq35}) and (\ref{eq36}) and using the definition
of $\lambda = \ell (\ell + 1)$, we obtain the $\ell$ values as
\begin{equation}
\label{eq38}
\ell = \sqrt {\frac{m^2 + \beta + \sqrt {(m^2 + \beta )^2 - \gamma ^2} }{2}}
+ n
\end{equation}
If we insert $\ell$ values obtained by Eq.(\ref{eq38}) into
eigenvalues of radial part of the Klein-Gordon equation with
non-central potential given by Eq.(\ref{eq29}), we finally find the
energy eigenvalues for a bound electron in the presence of a
non-central potential by Eq.(\ref{eq2})
\begin{equation} \label{eq39}
E_{Nnm} = M\frac{\left[ {\left( {N + \sqrt {\frac{m^2 + \beta +
\sqrt {(m^2 + \beta )^2 - \gamma ^2} }{2}} n + 1} \right)^2 -
\frac{\alpha ^2}{4}} \right]}{\left[ {\left( {N + \sqrt {\frac{m^2 +
\beta + \sqrt {(m^2 + \beta )^2 - \gamma ^2} }{2}} + n + 1}
\right)^2 + \frac{\alpha ^2}{4}} \right]}
\end{equation}
where $\beta = (\varepsilon + M)\beta $ and $\gamma = (\varepsilon +
M)\gamma $. The non-relativistic limit ($\alpha \ll 1$) of the
energy spectrum for the Hartmann problem where $\beta \ne $0 and
$\gamma $=0 is,
\begin{equation}
\label{eq40}
E_{Nnm} = - \frac{M\alpha ^2}{2}\left( {N + n + 1 + \sqrt {m^2 + \beta
\left( {\varepsilon + M} \right)} } \right)
\end{equation}
Then, the wave functions of polar-angle part of the Klein-Gordon
equation, using $\sigma $ and $\pi $ in Eqs.(\ref{eq13}) to
(\ref{eq15}), are obtained:
\begin{equation}
\label{eq41}
\phi = \left( {1 - x} \right)^{B + C / 2}\left( {1 + x} \right)^{B - C / 2}
\end{equation}
\begin{equation}
\label{eq42}
\rho = \left( {1 - x^2} \right)^B\left( {\frac{1 + x}{1 - x}} \right)^{ -
C}
\end{equation}
\begin{equation}
\label{eq43}
y_n = B_n \left( {1 - x} \right)^{ - (B + C)}\left( {1 + x} \right)^{ - (B -
C)}\frac{d^n}{dx^n}\left[ {\left( {1 + x} \right)^{n + B - C}\left( {1 - x}
\right)^{n + B + C}} \right]
\end{equation}
where $B = \sqrt {\frac{m^2 + \beta + u}{2}} $ and $C = \sqrt
{\frac{m^2 + \beta - u}{2}} $. The polynomial solution of $y_n$ is
expressed in terms of Jacobi polynomials which are one of the
ortogonal polynomials, giving $ \approx P_n^{(B + C,{\kern 1pt}
{\kern 1pt} B - C)} (x)$. Substituting Eqs.(\ref{eq41}) to
(\ref{eq43}) into Eq.(\ref{eq12}), the corresponding wave functions
are found to be
\begin{equation}
\label{eq44}
\Theta _n (x) = N_n \left( {1 - x} \right)^{(B + C) / 2}\left( {1 + x}
\right)^{(B - C) / 2}P_n^{(B + C,{\kern 1pt} {\kern 1pt} B - C)} (x)
\end{equation}
where $N_n $ is normalization constant determined by $\int\limits_{
- 1}^{ + 1} {\left[ {\Theta _{n}(x)} \right]^{{\kern 1pt} {\kern
1pt} {\kern 1pt} 2}dx} = 1$ and using the orthogonality relation of
Jacobi polynomials \cite{14,15}, the normalization constant becomes
\begin{equation}
\label{eq45}
N_n = \sqrt {\frac{(2n + 2B + 1)\Gamma (n + 1)\Gamma (n + 2B + 1)}{2^{2B +
1}\Gamma (n + B + C + 1)\Gamma (n + B - C + 1)}}
\end{equation}
\section{Conclusions}\label{sec6}
This paper presented a different approach, the Nikiforov-Uvarov (NU)
method, to the calculation of the non-zero angular momentum
solutions of the relativistic Klein-Gordon equation. Exact
eigenvalues and eigenfunction for the Klein-Gordon equation in the
presence of the non-central equal scalar and vector potentials are
derived easily. In the non-relativistic limit, the energy eigenvalue
spectrum is shown to be equivalent to Hartmann one and the radial
and polar angle wave functions are found in terms of Laguerre and
Jacobi polynomials respectively. The method presented in this study
is general and worth extending to the solution of other interaction
problems.
\section*{Acknowledgments}
This paper is an output of the project supported by the Scientific
and Technical Research Council of Turkey (T\"{U}B\.{I}TAK), under
the project number TBAG-2398 and Erciyes University (FBA-03-27,
FBT-04-15, FBT-04-16). The authors wish to thank Professor
Co\c{s}kun \"{O}nem for many helpful discussions and suggested
improvements to the paper.

\end{document}